# Stimulated Emission from 2D CdSe/CdS Nanoplatelets Integrated in a Liquid-Core Fiber

*Veronika Adolfs[1,2], Dominik A. Rudolph[2,3,4], Simon Spelthann[1,2,†,*], Artsiom Antanovich[3,4], Dan H. Chau[1], Mario Chemnitz[5,6], Markus A. Schmidt[5,7], Jannika Lauth[2,3,4,8,*], Michael Steinke[1,2,9,*]*

[1]Leibniz University Hannover, Institute of Quantum Optics, Welfengarten 1, D-30167 Hannover, Germany

[2]Cluster of Excellence PhoenixD (Photonics, Optics, and Engineering – Innovation Across Disciplines), Welfengarten 1A, D-30167 Hannover, Germany

[3]Leibniz University Hannover, Institute of Physical Chemistry and Electrochemistry, Callinstraße 3A, D-30167 Hannover, Germany

[4]Leibniz University Hannover, Laboratory of Nano and Quantum Engineering (LNQE), Schneiderberg 39, D-30167 Hannover, Germany

[5]Leibniz Institute of Photonic Technology, Albert-Einstein-Straße 9, D-07745 Jena, Germany

[6]Institute of Applied Optics and Biophysics, Philosophenweg 7, D-07743 Jena, Germany

[7]Otto Schott Institute of Material Research, Fraunhoferstraße 6, D-07745 Jena, Germany

[8]University Tübingen, Institute of Physical and Theoretical Chemistry, Auf der Morgenstelle 18, D-72076 Tübingen, Germany

[9]Leibniz University Hannover, QUEST-Leibniz-Research School, Callinstraße 36, D-30167 Hannover, Germany

ABSTRACT: Colloidal nanocrystals are unique optical gain materials due to their high intrinsic absorption, excellent quantum yield, and tunable emission. However, integration of colloidal nanocrystal solutions into photonic systems for lasing applications is challenging since high concentration levels are required for optical amplification. Here, we address this challenge by integrating colloidally dispersed core/crown CdSe/CdS 2D nanoplatelets in liquid-core optical fibers as a scalable platform. The platform allowed achieving sufficiently effective gain for amplified spontaneous emission at a threshold as low as 1.8 kW/cm$^2$ under quasi-CW pumping, even at a concentration two orders of magnitude lower than the minimal concentration considered to be required for gain in conventional colloidal quantum dots. We show that the low-loss optical waveguiding of the fiber is crucial for efficient stimulated emission, rendering liquid-core fibers as a promising and unique platform to realize lasers based on colloidally dispersed nanocrystals.

Colloidal nanocrystalline emitters represent a promising optical gain medium and have been studied in various configurations and material compositions.[1–3] Among these emitters, 2D semiconductor nanoplatelets (NPLs) stand out because they exhibit high intrinsic absorption cross-sections,[4,5] narrow emissions ranging from the ultraviolet to the near-infrared,[6–11] and low thresholds for amplified spontaneous emission (ASE) and lasing.[12–21] CdSe NPLs with a CdS crown constitute a comparably simple heterostructure with emission wavelengths tunable via the number of atomic layers.[22–24] The core/crown design leads to high photoluminescence (PL) quantum yields since nonradiative recombination at edge defect states is efficiently suppressed.[22–25]

To date, ASE or lasing with semiconductor nanomaterials has mainly been realized in solid thin films, due to ease of preparation and high packing densities typically required for gain.[12–20,26–28] However, such films can be prone to various loss channels such as scattering or Förster resonance energy transfer and subsequent charge carrier trapping.[29] These challenges can be alleviated by using dispersed nanocrystals in a liquid environment. Corresponding optofluidic setups also offer further advantages, including a relatively straightforward exchange of the gain medium, to either replenish it to prevent degradation[30] or to replace it altogether to cover different emission wavelengths.[13] Despite the advantages, achieving optical gain from colloidal solution is challenging mainly due to the limited nanocrystal solubility. Park et al. estimated that to enable gain with a colloidal quantum dot solution, a semiconductor volume fraction of 2% is necessary, just to overcome inherent nonradiative Auger recombination losses.[31] Further, since the nanomaterial concentration required for gain must also compensate optical losses incurred by the photonic environment, either specific solvents with high nanocrystal solubility must be used,[32] or the optical device itself has to provide a low-loss environment. Consequently, only few studies have been published reporting ASE or lasing in colloidal nanomaterial solutions employing cuvettes or short capillaries.[21,32–35] This demonstrates that, while the maturity of the NPL design and synthesis is quite evolved and well-controlled, scalable low-loss platforms for their optical integration in solution are missing to enable their broader deployment as a gain medium in optical technologies.

As an innovative approach to enhance the effective NPL gain even at low concentrations, we propose a scalable platform for the optical integration of colloidal nanocrystals such as NPLs. Particularly, we utilize optical fibers, which efficiently collect the NPL emission and confine it on a small cross-section but are easily scalable in the longitudinal direction due to low-loss waveguiding. Fibers based on fused (vitreous) silica offer excellent technological maturity and chemical, thermal, and mechanical robustness.[36] Such fibers form the basis for modern data communication[37] and active fibers doped with trivalent lanthanide ions enabled kW-class continuous wave (CW) or high-energy pulsed laser systems.[38] However, pronounced multiphonon quenching of fused silica prevents visible lasing from lanthanide-doped fibers[39] and, in turn, makes nanocrystals a promising wavelength-tunable laser gain medium. While direct doping of colloidal nanocrystals into fused silica cannot be achieved at the extreme temperatures above 2000 °C required for fiber fabrication,[40] liquid-core fibers (LCFs) overcome their solid counterparts' disadvantage and open a promising path for the photonic integration of colloidal nanocrystals into optical systems.[41] These fibers have recently been receiving substantial interest particularly in nonlinear photonics due to unique properties such as non-instantaneous nonlinear response[42] or the ability to tune dispersion in real-time through temperature modulation.[43] Specifically, nanomaterial dispersions can be straightforwardly introduced into LCFs under ambient conditions via capillary action. Recently, Zhang et al. followed this idea by demonstrating gain from quantum dots embedded in an LCF.[44] However, due to the high competition between radiative emission and non-radiative Auger recombination in quantum dots, a complex ternary-alloyed CdZnSe/ZnSeS/ZnS structure with a precisely engineered core/shell interface and concentrations in the range of hundreds of grams per liter were required to achieve this result. NPLs, on the other hand, offer practical advantages including a low inherent ASE threshold and low Auger recombination rates arising from their 2D structure and the strong confinement.[28,45]

Here, for the first time, we integrate core-crown NPLs in LCFs and demonstrate ASE (synonymously referred to as stimulated emission) under quasi-CW pumping. Using a numerical spectral decomposition, we find that optical gain can only build up for red-shifted biexcitons, an observation we put into context with two common explanations of the gain mechanism in NPLs. We highlight that the low-loss waveguiding of the LCFs is a unique approach to obtain gain, even at comparably low NPL concentrations. The impact of our results is two-fold:

1) Our approach opens a promising path to significantly extending the spectral regime of fiber lasers towards the visible wavelength range.
2) LCFs offer a scalable, chemically inert, and environmentally robust platform for photonic integration of NPLs and further colloidal nanocrystalline emitters.

Within this work, we used 4.5 monolayer thick CdSe/CdS core/crown NPLs at a low concentration of ~1.56 µmol/L (~1.02 g/L; volume fraction of ~0.015%) prepared according to published procedures (see SI).[6,22,23,46] The NPLs have a rectangular shape with an average lateral size of 27 × 9 nm$^2$ (see **Figure 1a and S1a)**, starting from an average CdSe core size of 14 × 4 nm$^2$ (see **Figure S1b**). **Figure 1b** shows the absorption and the PL of the NPLs as measured in a cuvette. The latter is centered at 2.41 eV (~ 515 nm) and features a quantum yield of 85 %. The NPLs were dissolved in tetrachloroethylene (TCE), which was used due to its high boiling point (121 °C), high transparency, and high refractive index of 1.51 (at 530 nm),[47] which enables waveguiding within LCFs via total internal reflection. **Figure 1c** shows the evolution of the recorded NPL emission spectra from an LCF with a 26 µm core. In direct comparison with the excitonic (monocomponent) PL of the NPLs, we find a pronounced additional red-shifted (30-45 meV) emission. It indicates the formation of biexcitons and, potentially, ASE as it has been observed for similar NPL systems.[12,48]

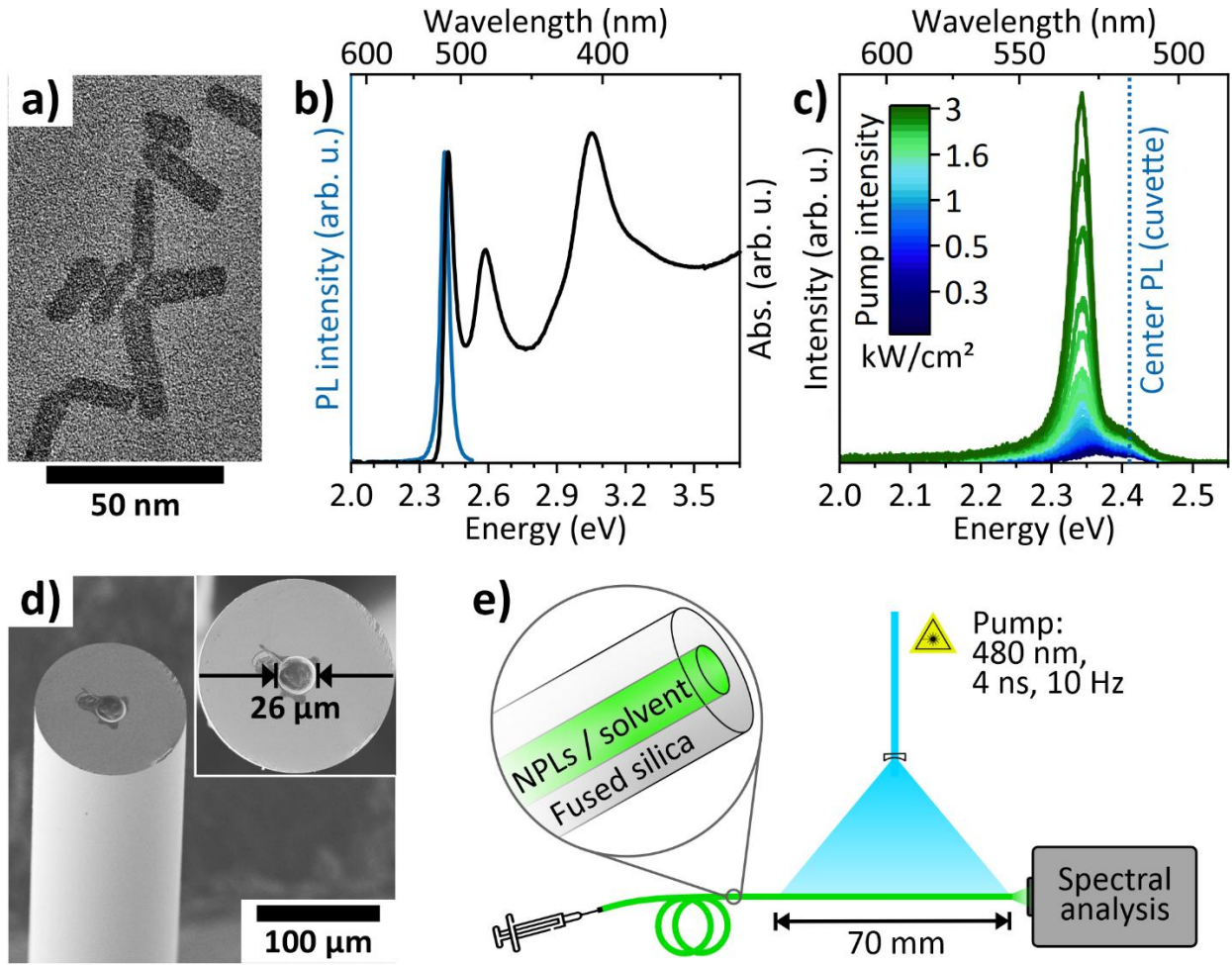

**Figure 1:** a) TEM image of the colloidal 2D CdSe/CdS core-crown NPLs. b) Absorbance (black) and PL (blue) of the NPLs in TCE. c) Emission spectra obtained from an LCF filled with a colloidal solution of CdSe/CdS NPLs under an increasing intensity of the quasi-CW pump laser. d) SEM image of the LCF used in the experiments. e) The scheme of the setup used to laterally pump the fiber by a quasi-CW laser system.

An SEM picture of the LCF and a schematic sketch of the setup used to obtain our results is presented in **Figure 1d** and **e** (a full description of the setup is provided in the SI). We transversally pumped a 7 cm long section of the LCF by an optical parametric oscillator tuned to 480 nm. The pump laser provides 4 ns pulses with an energy density of up to 21 µJ/cm², which corresponds to a maximum (peak) intensity of 5.4 kW/cm$^2$. Note that the pump pulses are five orders of magnitude longer than fs-lasers that are commonly used to pump the short-lived excited NPL states.[12,13,15,27] Since the expected biexciton (gain) lifetime lies in the order of a few hundred ps[49,50] – about ten times shorter than our pulse duration – our experiments are operated in the quasi-CW regime.

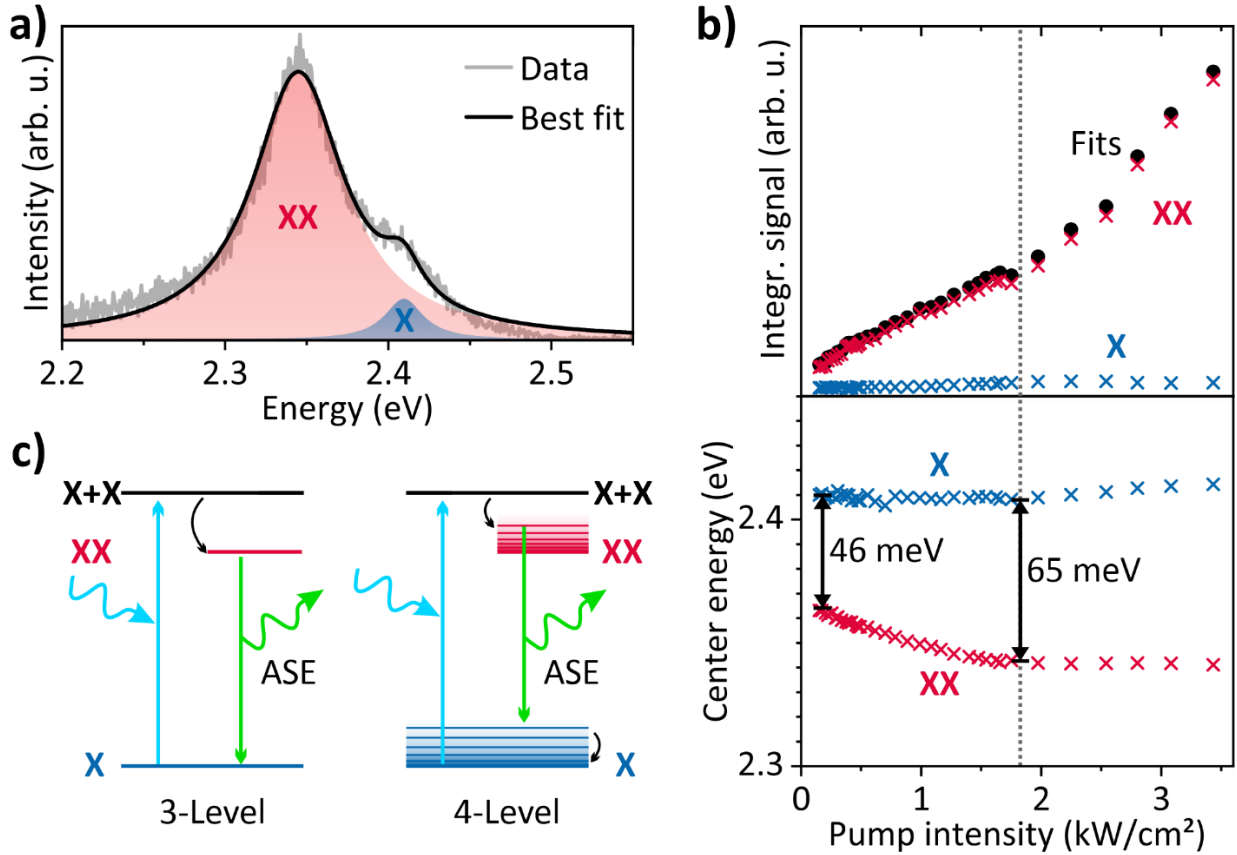

**Figure 2.** a) Exemplary decomposition of the recorded emission spectrum (from the LCF) at a pump intensity of ~1.8 kW/cm$^2$ into two spectral components, showing an excitonic (X) and biexcitonic (XX) contribution. b) Evolution of the spectrally integrated emissions (top) and the center energies of the fitted X and XX components (bottom). The dashed line indicates an ASE threshold of 1.8 kW/cm$^2$. c) Scheme of the relevant NPL states involved in the optical amplification process in a three-level (left) and four-level system (right).

To analyze the measured spectra and confirm the onset of ASE, we numerically decomposed the spectra by a model consisting of a Lorentz profile for the excitonic (X) and a Pseudo-Voigt profile for the biexcitonic (XX) component. **Figure 2a** shows a representative result of our fitting procedure. For an increasing pump intensity, the integrated signals (slopes) of the two fitted components and their center energies reveal two operation regimes, see **Figure 2b**. The first regime corresponds to pump intensities below 1.8 kW/cm$^2$, where the slopes increase linearly at moderate rates corresponding to spontaneous excitonic and biexcitonic emission. The second regime corresponds to pump intensities above 1.8 kW/cm$^2$, where the biexcitonic slope is significantly steeper than before, indicating ASE. The threshold of 1.8 kW/cm$^2$ corresponds to a energy density of 7.2 µJ/cm$^2$, which is significantly lower than previous results of dispersed nanocrystals (1 mJ/cm$^2$ for quantum dots in an LCF from Zhang et al.,[44] 30 µJ/cm$^2$ for red emitting and 44 µJ/cm$^2$ for green emitting NPLs in short capillary tubes from Delikanli et al.[21]). Note that this comparison is rather qualitative as those previous results were achieved with fs-pumping, whereas we use quasi-CW pumping. The ASE threshold can also be identified in the evolution of the excitonic and biexcitonic center energies, see **Figure 2c**. At the lowest pump power level, the excitonic and biexcitonic center energies are separated by ~46 meV. Towards the threshold, the biexcitonic center energy shifts further into the red until it remains constant at 2.34 eV (~530 nm) above the threshold, with a separation of ~65 meV with respect to the excitonic emission. This means, the ASE emerges on the red tail of the spontaneous biexcitonic emission (see further fitting results in the SI).

The red shift of the ASE gain can be understood when considering two common but complementary explanations of how gain in NPLs occurs.[51,52] From a laser physics perspective, gain originates from either a three-level system or a four-level system, as shown in **Figure 2d**. In the three-level picture, the requirement for gain is an average biexciton density $\langle N_{xx} \rangle$ that is larger than the exciton density $\langle N_x \rangle$. This explanation is directly adapted from the common understanding of the processes in quantum dots, where gain is achieved at an exciton density $\langle N_x \rangle$ greater than 1.15 and where two excitons are assumed to always form a biexciton.[51] However, this picture does not allow for an energy shift of the stimulated emission with respect to the spontaneous biexcitonic emission and cannot explain our observation of a red-shifted ASE. Such a red shift occurs, however, if the 2D mobility of the excitons and biexcitons and their corresponding kinetic energy is considered, as has been demonstrated by Geiregat *et al.*[52] Here, the authors argued 1) that biexcitons and excitons on NPLs exist in a thermodynamic equilibrium and 2) that thermal occupation (following a Boltzmann distribution) of the kinetic energy states essentially provides a 4-level system. In combination with momentum conservation, the thermally occupied states yield gain at lower energies (i.e. red-shifted) than the spontaneous biexcitonic emission. Indeed, such red-shifted gain occurs at densities $\langle N_{xx} \rangle$ less than $\langle N_x \rangle$ and is in line with our observations.

The recorded spectra and the fitted slopes (see **Figure 2b**), however, imply that there are significantly more biexcitons than excitons, i.e., that $\langle N_x \rangle$ is actually less than $\langle N_{xx} \rangle$. This seems to be in contradiction with the explanation provided by Geiregat *et al.* but can be resolved by the following two arguments. 1) Due to the quasi-CW pumping scheme, we cannot reliably anticipate the expected exciton and biexciton densities. However, additional calculations in the SI demonstrate that $\langle N_{xx} \rangle$ is expected to be less than $\langle N_x \rangle$ even assuming a best-case scenario. This scenario corresponds to an infinite charge carrier lifetime, i.e., charge carriers accumulate during the pump pulse without decay. 2) We provide an additional analysis in the SI (see **Figure S9**), indicating high kinetic energies of the biexcitons that correspond to effective temperatures up to 500 K above room temperature. Assuming that excitons will experience similar temperatures, a high number of them are expected to be in momentum-forbidden dark states, i.e., the quantum yield of their emission will be significantly reduced. In contrast, such momentum-forbidden dark states do not exist for biexcitons. The combination of these two arguments explains the seemingly high relative number of biexcitons implied by the slopes of the fitted excitonic and biexcitonic emissions (see **Figure 2b**).

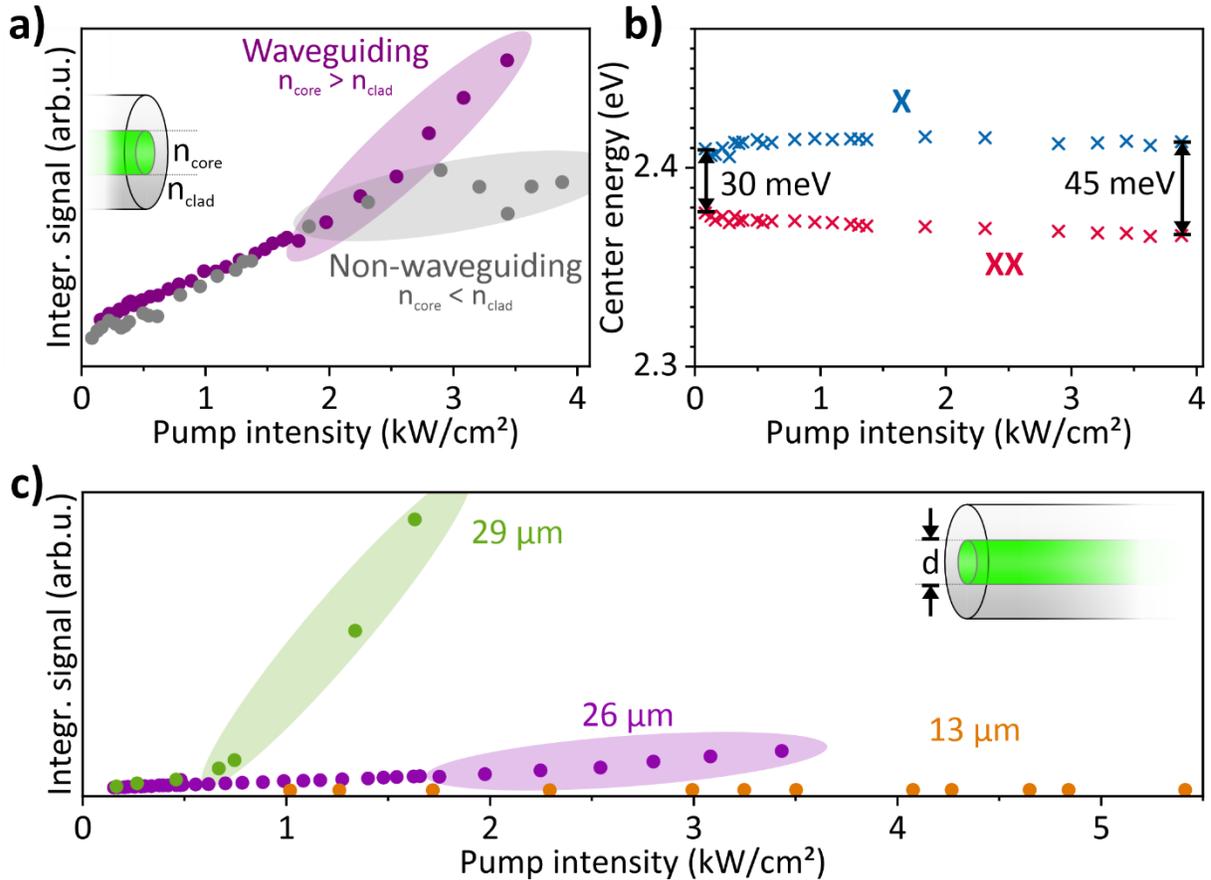

**Figure 3.** a) Integral emission intensity in LCFs filled with CdSe/CdS NPLs dispersed in hexane (no waveguiding) and TCE (waveguiding). b) Evolution of the center energies of the fitted excitonic and biexcitonic components for the hexane-filled fiber. c) The slopes obtained from the LCFs with different inner diameters of 13 μm (orange), 26 μm (purple), and 29 μm (green) demonstrating the unique scaling properties provided by LCFs.

Since the photonic environment provided by the LCF is crucial for the generation of stimulated emissions, we performed further investigations of the LCF parameters. First, we conducted experiments with hexane NPL dispersions to suppress the waveguiding of the fiber, since the refractive index of hexane is 1.38 at 530 nm,[53] which is below that of fused silica (1.46[54]). For brevity, we will refer to the LCF filled with hexane as a non-waveguiding fiber to distinguish it from the waveguiding fibers filled with NPLs dispersed in TCE. The slope of the non-waveguiding fiber, shown in **Figure 3a**, obtained by integrating the recorded spectra, coincides well with the corresponding slope of the waveguiding fiber up to its ASE threshold (1.8 kW/cm$^2$). This is expected because even for the waveguiding fiber, spontaneous emission originates primarily from the fiber tip because any such emission upstream of the fiber is effectively reabsorbed along its length. Only at the ASE threshold, the emission in the waveguiding fiber is sufficiently red-shifted to reach stimulated emission. However, the slope obtained from the non-waveguiding fiber lacks indication of stimulated emission above the threshold of the waveguiding fiber. It even rolls off at the highest pump intensities, which we attribute to an increasing number of excitons in momentum-forbidden dark states (see discussion above). We numerically decomposed the spectra obtained from the non-waveguiding fiber into two components (X and XX) and show their corresponding center energies in **Figure 3b** (**section S5.3** in the SI). The lack of a threshold is also supported by the comparably low excitonic to biexcitonic energy separation of 30 meV (lowest pump intensity) and 45 meV (highest pump intensity), which does not increase as significantly as for the waveguiding fiber (see **Figure 2c** for comparison).

Even with a waveguiding fiber, ASE only occurs if the effective (single pass) gain provided by the LCF is sufficiently high to overcome the losses of the system, which is not always the case at the NPL concentration we employed. We repeated our initial experiment using a waveguiding, i.e., TCE-filled, LCF with a smaller core diameter of 13 μm (see **Figure 3c**). We did not observe any indication of ASE for this core size in the range of the used pump intensities, certainly because the absolute number of NPLs and their corresponding effective gain was too small to overcome the losses. The three main parameters to tune the threshold and scale the slope efficiency are the LCF core diameter, the fiber loss, and the NPL concentration. The latter has been kept constant at 1.56 µmol/L for all experiments. To showcase the influence of the fiber diameter, we repeated our experiment using a waveguiding LCF with 29 µm core diameter and compare the slopes with the former experiment using the 26 µm core LCF in **Figure 3c**. Strikingly, the ASE threshold decreases significantly by around 60 % to a pump intensity of 0.7 kW/cm$^2$. In parallel, the slope efficiency increases by a factor of around 18. This drastic improvement in the performance parameters of our system can be attributed to an 11.5 % increase in the LCF core cross-section, resulting in a 25 % increase in the absolute number of NPLs, i.e., effective gain per unit fiber length. Additionally, we determined the different background losses of both LCFs (26 and 29 µm core, see SI). Due to differences in the fiber fabrication process, the 29 µm LCF features a waveguide loss of 0.005 dB/cm, which is one order of magnitude lower than the loss of the 26 µm LCF (0.05 dB/cm). Indeed, besides the larger cross section, the lower loss also positively influences the threshold and slope efficiency. The numerical decomposition analysis for the 29 µm LCF shows (see **Figure S10** in the SI for the center energies) that the ASE threshold is reached once the biexcitonic center energy is ~2.34 eV, which is the same as for the 26 µm LCF. In addition, the ASE is separated by 59 meV (65 meV for the 26 µm LCF) from the excitonic emission at the threshold. These results indicate a similar role of the excitons and biexcitons in the gain process, independent of the LCF parameters. However, future investigations will be employed to reveal the correlations of the gain process with the LCF parameters.

In conclusion, we demonstrate a well-implemented, robust and scalable platform for the photonic integration of nanocrystalline emitters such as 2D NPLs in solution. NPLs are a unique and promising gain material with great potential for optofluidic applications, which to date remain largely unrealized due to performance limitations arising from the inherent low emitter concentration. In this work, we overcome this obstacle by integrating CdSe/CdS core/crown NPLs in waveguiding LCFs for the first time. Even at a comparatively low concentration of 1.56 µmol/L and under quasi-CW pumping, we observe ASE above a low threshold of 1.8 kW/cm$^2$. To further contextualize our findings, we perform numerical decompositions of the recorded emission spectra, which reveal the role of the excitons and biexcitons. Our analysis aligns well with the NPL gain model by Geiregat et al.[52] Furthermore, we have investigated the impact of the photonic environment on the ASE enabled by altering the LCF core size. We show that thresholds and slope efficiencies can be positively affected by increasing the fiber diameter and decreasing the fiber loss. The parameters of the LCFs (length, loss, core diameter) can readily be scaled and adjusted further towards CW pumping with affordable diode lasers – a circumstance that was part of the significant success of lanthanide-doped fiber lasers. LCFs can also be fusion spliced and combined with standard fiber technology[41,55] and adapted to be operated with various nanoemitter classes. Thus, LCFs will certainly be an impactful contribution to the challenge of bringing colloidal nanomaterials into lasing applications. In addition, NPL-filled LCFs constitute a promising path towards highly efficient and wavelength-tunable fiber lasers in the visible spectral regime.

## ASSOCIATED CONTENT

**Supporting Information**. The SI is free of charge and available as a PDF file. It contains the following additional information: synthesis of CdSe/CdS core/crown NPLs; additional TEM images used to determine NPL dimensions and size histograms of CdSe core-only and CdSe/CdS core/crown NPLs; PL spectra of CdSe/CdS core/crown NPLs in TCE and hexane; Absorption spectra of CdSe core-only and CdSe/CdS core/crown NPLs; calculation of exciton binding energy for CdSe/CdS core/crown NPLs; calculation of CdSe/CdS core/crown NPLs concentration; fabrication of fused silica fibers; characterization of fibers by SEM; discussion of numerical aperture and V-number for the used LCFs; characterization of fiber loss parameters; details of the optical pump setup; emission spectra from CdSe/CdS core/crown NPLs from all LCFs (13, 26 and 29 µm) in their respective solvents; discussion and characterization of the ratio between excitons and biexcitons of CdSe/CdS core/crown NPLs in LCFs; additional characterization of CdSe/CdS core/crown NPLs in 29 µm TCE-filled and 26 µm hexane-filled LCFs.

AUTHOR INFORMATION

**Corresponding Authors**

Simon Spelthann, Michael Steinke and Jannika Lauth

**Present Addresses**

†Ruhr-University Bochum, Simply Complex Lab, Universitätsstraße 150, D-44801 Bochum, Germany

**Author Contributions**

V. A. and D. R. contributed equally to this work. Conceptualization: M. S, S. S., and J. L.; synthesis and characterization of the nanoplatelets: D. R.; liquid-core fiber experiments: V. A.; fiber production and LCF expertise: M. C. and M. A. S.; visualization and writing – original draft: V. A. and D. R.; writing – review and editing: all authors; supervision: J. L. and M. S.; resources: J. L. and M. S. All authors reviewed the manuscript and approved the final version.

**Notes**

The authors declare no competing financial interest.

ACKNOWLEDGMENT

V. A., D. R., S. S., M. S. and J. L. gratefully acknowledge funding by the Deutsche Forschungsgemeinschaft (DFG, German Research Foundation) under Germany's Excellence Strategy within the Cluster of Excellence PhoenixD (EXC 2122, Project ID 390833453). M. C. acknowledges funding by the Carl-Zeiss Foundation through the NEXUS program (project P2021-05-025). We thank Ronja Stephan and Katharina Hausmann for drawing the 13 μm and 26 μm fibers and the colleagues of M. A. S. and M. C. at IPHT Jena for the 29 μm hollow core fiber. We are also grateful to Frank Steinbach for performing the SEM measurements on the fiber. J. L. is thankful for funding by the Ministry for Science and Culture of the State of Lower Saxony (MWK) for a Stay Inspired: European Excellence for Lower Saxony Grant (Stay-3/22-7633/2022) and for additional funding by an Athene Grant of the University of Tübingen (by the Federal Ministry of Education and Research (BMBF) and the Baden-Württemberg Ministry of Science as part of the Excellence Strategy of the German Federal and State Governments).

ABBREVIATIONS

NPL, nanoplatelets; ASE, amplified spontaneous emission; PL, photoluminescence; CW, continuous wave; LCF, liquid-core fiber; TCE, tetrachloroethylene